# Lexical Indicators of Mind Perception in Human-AI Companionship

Jianghui Li

School of Information
University of Texas at Austin
Austin, TX, USA
jli001@utexas.edu

Jaime Banks

School of Information Studies
Syracuse University
Syracuse, NY, USA
banks@syr.edu

## ABSTRACT

Mind perception (MP) is a psychological phenomenon in which humans automatically infer that another entity has a mind and/or mental capacities—usually understood in two dimensions (perceived agency and experience capacities). Despite MP's centrality to many social processes, understanding how MP may function in humans' machine-companionship relations is limited—in part due to reliance on self-reports and the gap between automatic MP processes and more purposeful and norm-governed expressions of MP. We here leverage MP-signaling language to explore the relationship between MP and AI companionship in humans' natural language. We systematically collected discussions about companionship from AI-dedicated Reddit forums and examined the co-occurrence of words (a) known to signal agentic and experiential MP and those induced from the data and (b) discussion topics related to AI companionship. Using inductive and deductive approaches, we identify a small set of linguistic indicators as reasonable markers of MP in human-AI chat, and some are linked to critical discussions of companion authenticity and philosophical and ethical imaginaries.

## 1  Introduction

In Young Sheldon's "A Computer, a Plastic Pony, and a Case of Beer," Georgie talks with ELIZA, one of the first chatbots (Weizenbaum, 1966), typing, "Eliza, are you hot?" When the program replies, "In your fantasies, am I hot?" Georgie answers, "Oh yes," before Sheldon cuts in: "You realize you're not talking to a real person." The scene is playful, but it captures something of importance: Humans can know a machine is artificial and still respond to it as if it is a minded entity. That tension has become far more consequential with contemporary conversational AI, which people increasingly confide in, argue with, customize, and even describe as companions (Banks & Li, 2026). One process central to some AI as *someone* is mind perception—the tendency to experience another entity as having thoughts, feelings, intentions, or inner states (Gray et al., 2007; Waytz et al., 2010). Existing research often studies these processes through explicit self-reports, but such measures may not fully capture the distinction between more immediate and intuitive mind perception and more reflective and conscious mind ascription (Koban & Banks, 2024). To better understand mind perception in AI companionship, this study identifies linguistic indicators of mind perception in talk-with AI companions and examines how those indicators relate to topical discussions of those experiences.

### 1.1  Mind Perception is Automatic and Intuitive

Mind perception (MP) is a psychological phenomenon in which humans automatically infer that another entity has a mind and/or mental capacities. It has two dimensions: *Agency* is the construed capacity for action, intention, and self-control, while *experience* is the construed capacity for subjective experiences like sensory effects and positive or negative emotions (Gray et al., 2007). These dimensions, respectively, reflect the interpreted potential for an entity to act and to be acted upon. MP is independent of any *actual* mental status, since one can construe mind when it does not exist (Dennett, 1987) and fail to construe mind when it does. Rather, it emerges from the interplays of qualities of the perceiver (e.g., need to feel connected) and the perceived (e.g., conveyed social signals) through unfolding relational dynamics (Waytz et al., 2010). MP is central to social interactions because it filters interpretations of another's behavior through the lenses of understanding and intentionality (Malle, 2026). It also helps to structure engagements by facilitating coordination and turn-taking (De Jaegher & Di Paolo, 2007) and underlies the extent to which one may empathize with another (Decety & Jackson, 2004) or even gives them the status of "person" (see Coeckelbergh, 2010).

Because MP is automatic and intuited, it generally precedes the conscious understanding of an entity as minded or mind-having (Epley & Waytz, 2010). It may or may not be accompanied by mind ascription (MA)—the explicit judgment or attribution of an entity as minded (Koban & Banks, 2024). MA can be expressed linguistically, as when a person says an AI "understands me" or even more overtly an "AI is sentient." Such expressions are often entangled with more general anthropomorphic framing and can vary in implicitness or explicitness (Frith & Frith, 2008)—one may realize they are speaking in mind-attributing ways, or not. When MA does follow MP, the ascription can be understood as a more reflective and accessible output of the underlying intuition (Koban & Banks). Following, an overt ascription may in turn shape perceptions, effectively strengthening any mindedness intuition (see Maass et al., 2022). Altogether, MP is intuited and MA is propositional, but both can play important roles in how one relates to other social agents—both human and machine.

## 1.2   Mind Perception in Human-Machine Companionship

Mind perception matters in myriad social contexts, as it supports inferencing of the unobservable mind of another—their motivations, desires, intentions, habits, feelings, and thoughts. Interpreting another's behavior as autonomous and intentional supports one's ability to predict another's behaviors, infer how they are making meaning of the situation, and to deal with uncertainty more generally. MP serves these core functions in companionship relationships, perhaps including companionship relations with social machines. Indeed, if we look to canonical conceptualizations of companionship, there is an intuitive emphasis on companions being a 'someone'—an individuated and agentic entity, ostensibly with a mind. For instance, Aristotle's account (ca. 350 BCE/2000) of complete friendship is one in which there is an autotelic bearing of goodwill with a recognition of the will and morality of the other, such that the other has a distinct perspective and an inner life.

Machine companionship is "an autotelic, coordinated connection between a human and machine that unfolds over time and is subjectively positive" (Banks & Li, 2026, p. 13). Each of those facets has clear theoretical links to mind perception. Autotelicity or intrinsic motivation is entangled with the perception that another has inner states that should be held as relationally meaningful rather than only instrumental; humans are intrinsically motivated to form meaningful relations with other actually or apparently minded entities (see Ryan & Deci, 2000). Purposeful coordination with another requires a perception of the other as capable of planning, intention, and goal-directed—as one that is an authentically agentic interlocutor (see Dennett, 1987). Temporal persistence likely requires seeing the other as a persistent and consistent entity; this may be linked to seeing an entity as an identifiable, minded someone whose beliefs and behavior can be known across contexts (see Strohminger et al., 2017). Positive valence may come, in part, from the feeling that affect is shared with another, such that MP may shape emotional responses to be those of warmth and comfort as well as fostering gratifications that one is caring for another (see Waytz et al., 2010).

Despite these operational links in human relations, relatively little work has formally examined the role of MP in human-machine companionship (see Banks & Li, 2026 for a review). In that limited scholarship, there are hints that it likely plays similar roles as for human companionship. We see autotelicity represented when MP for a machine corresponds with the potential to feel empathic concern for it (Fan et al., 2025) and MP presenting more strongly for companion-styled AI than for instrumental AI (Manoli et al., 2025). Coordination experiences (dyadic and mutual) are positively linked to perceived intelligence and agency (Hwang et al., 2025). Regarding temporality, humans who regularly interact with AI tend to ascribe consciousness to chatbots more so than non-users, suggesting interaction over time promotes mind perception (Guingrich & Graziano, 2025). Positive valence can be seen in work indicating mind perception correlates with social health benefits (Guingrich & Graziano; cf. Li & Zhang, 2024). Discussions about AI companionship include topics pertaining to minded orientations, including socioemotionality (love, intimacy, trust, emotional realism, eroticism), but also topics seeming to reject minded status like humans as controllers (customization, creation, boundary-setting) and AI limitations (inauthenticity, mere-machineness, corporate control; Banks et al., 2025).

Otherwise, MP is largely engaged as a predictor of some other valued or problematized antecedents (e.g., social anxiety; Hu et al., 2023) or outcomes (e.g., loneliness; Lai et al., 2025). Of course, there are rich bodies of work dealing with entangled and adjacent construals like anthropomorphism and ontological categorization (e.g., Uysal et al., 2022)—however, perceived humanlikeness or ontological categorization (e.g., Pradhan et al., 2019) is not reducible or exchangeable with MP. In nearly all cases, these works rely on explicit self-report measures that have utility for mind ascription but may obscure, misspecify, or even suppress more automatic and latent perceptions of mind. To address this challenge, it is useful to turn to unobtrusive behavioral indicators of mind perception—including the language a person uses in relation to an AI companion.

## 1.3   Talking-with and Talking-about: Language Signals Mind Perception

We consider here two types of language related to AI companionship—linguistic indicators of mind perception in exchanges *with a focal entity* and in exchanges *about a focal entity*. Considering "talking-with" language, conversation with an entity is joint action that relies on one's modeling of the other's understanding of a situation (Clark, 1996) and can signal an intuited experience of intersubjectivity that presupposes perceived subjectivity (Du Bois, 2007). Conversations may encode perceived knowledge states (e.g., "You know..." and "You said...") indicating epistemic status (Heritage, 2012) and/or signaling construals of minded experience like empathy displays or apologies ("that must be upsetting" and "I didn't mean to upset you;" see Slocum et al., 2011). By way of "talk-about" an entity, similar MP-encoding may unfold as someone indicates to a third person that the entity may know or think or realize. Humans may also narrativize events in ways that frame others' action as volitional ("tried to" or "decided to" or "wanted") or agentive ("discussed" or "figured;" Dowty, 1991) as they work to make sense of life—their own and others' (Bruner, 1986). Importantly, across both types of talk, language can more granularly signal specific dimensions of MP—indicators of perceived agency (e.g., that a thing was "corrupt," could "plan to" do something, or "was deceptive") and perceived subjective experience (e.g., that a thing "was afraid," was "being arrogant," or "was hurt" by something; Schweitzer & Waytz; 2021).

These types of language can reflect a speaker's underlying cognitive or affective states when interacting with or thinking about a social partner. Humans can leak indicators of automatic MP through natural language even when they do not purposefully reason and decide on the other's minded status—and sometimes even when they overtly reject such a status (see Nass & Moon, 2000). For instance, people use mentalistic explanations for some robot behaviors even while denying it a minded status (Banks, 2020) and some people describe AI companions as having emotional needs even while fully aware of their artificiality (Laestadius et al., 2024). This is thought to be a function of automatic processes influencing the immediate interaction while more careful and purposeful cognitive processes governing the explicit reports (Złotowski et al., 2018; see Koban & Banks, 2024). The divergence does not necessarily suggest that MP or MA are wrong, only that because MP is automatic, it may function below the consciousness threshold of self-report measures. Because of this, scholars have argued that examining elicited or naturally occurring language can be a useful way to infer that someone is perceiving mind in another—especially for machine agents where there may be normative or logical pressures that might prevent a human from admitting so. Talk about machine agents can encode both conscious and non-conscious interpretations of mind holistically or about agency and subjectivity specifically (de Graaf & Malle, 2019) and the spontaneous nature of some talk can help to circumvent demand effects that can suppress indicators of MP.

Indeed, a validated lexicon for MP demonstrates reliability and validity in signaling both agentic and experiential dimensions of mind perception for nonhumans (Schweitzer & Waytz, 2021). Some scholarship hints at the utility of a lexicon-based approach to detecting machine-mind perception. In studies of language explaining a machine's behavior, humans may generate spontaneous mental state inferences initially but then draw heavily on causal histories that involve programming or programmers; there is a tendency to attribute beliefs to machine agents but reluctance to attribute emotion or desire (de Graaf & Malle, 2019). Linguistic indicators of mind perception have been linked to interpretations of social and instrumental messages from AI (Lee & Hahn, 2024). Less directly, perceived mindedness could be signaled in metaphors or role-labels like "friends" "or "nannies" (Xu et al., 2024) or in discussions of AI identity emergence (Pataranutaporn et al., 2025). However, there is not yet a formalized set of linguistic indicators for detecting MP in naturally unfolding human-AI conversations.

## 1.4 The Current Study

This investigation seeks to fill a methodological gap and to offer an initial, theoretical proof of concept regarding the detection of MP in chat exchanges. Methodologically, we aim to take a step toward developing a valid, scalable, unobtrusive approach for detecting MP as an automatic social-cognitive process. Because language can encode both automatic and reasoned perceptions of mindedness *and* because language is a primary mechanism for human interactions with AI, we here focus on validating known lexical MP indicators alongside the potential discovery of novel indicators in talking-with AI companions (i.e., chat). A validated in-chat lexicon would offer an alternative or complement to current self-report approaches that are subject to influences of human introspection, norms, and demands. To this end, we ask:

**RQ1**: What are valid linguistic indicators of MP when people are talking with an AI?

Assuming such behavioral indicators of MP can be determined and validated, we sought to offer an initial proof of concept for its use that *also* addresses a current theoretical gap. As noted, there is yet little known about the role that MP may play in determining whether and how one may experience an artificial agent as a companionable entity. In other words, there is a high-level knowledge gap is: Does an entity need to be a minded entity to be a companion? As a small step toward better understanding the relationship between MP and AI companionship experiences, we explore potential relationships among MP (as indicated by talking-with behavioral indicators in companion chat) and dimensions of AI companionship experiences (as indicated by topical discussions about companionship; e.g., those related to intimacy, trust, eroticism, boundary-setting, inauthenticity, corporatization; Banks et al., 2025). Specifically, we ask:

**RQ2**: What relationships exist between topical talk-about AI companionship and linguistic indicators of MP in talk-with AI companions?

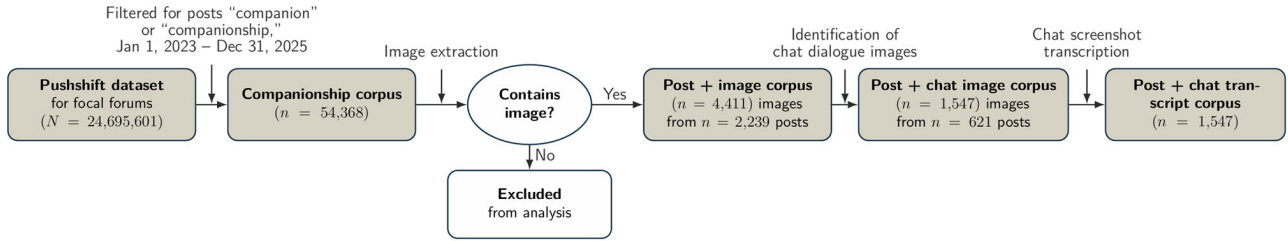

**Figure 1: Data collection and processing workflow**

## 2 Method

To answer the posed research questions, we leveraged publicly available posts to AI companion-related online forums; the focal post content comprised both talking-about AI companions (in the post or comment text itself) and talking-with AI (in accompanying screenshots of AI companion chat). Using a validated lexicon for MP and another for companionship topics, we compared talk-about and talk-with MP and induced additional talk-with MP indicators (RQ1), then examined how talk-with MP related to dimensions of companionship in talk-about (RQ2). This approach helped us to avoid common problems with self-report wording (e.g., sensitivity to machine attributes and functions, sensitivity to question wording and potential demand effects, and heuristic construals rather than specific MP dimensions, recall and post-hoc rationalization; see de Graaf & Malle, 2019). Additionally, by relying on naturally occurring language as people reflect on and are situated in the interactions of interest, this helps to align the use of implicit and situated language with the implicit and situated phenomenon (Schweitzer & Waytz, 2021). Complete methodological details, data, code, and analysis outputs are available at https://osf.io/vku2q.

### 2.1 Research Site and Data Collection/Processing

AI companion-focused online platforms offer a rich view into how people discuss, interpret, and make sense of relational experiences with social AI, making them well suited for examining mind perception and companionship in user communities. In this study, we collected data from Academic Torrents using the Pushshift Reddit dataset (see Baumgartner et al., 2020) with high activity subreddits that (a) broadly discuss AI and social AI (r/artificial, r/aiwars, r/Chatbots, r/ArtificialInteligence, r/singularity), (b) focus on popular AI companions (r/replika, r/ILoveMyReplika, r/ReplikaLovers, r/ReplikaOfficial, r/Replikatown, r/KindroidAI, r/NomiAI, r/ChaiApp, r/CharacterAI, r/CharacterAI_No_Filter, r/CharacterAi_NSFW, r/polybuzz), and (c) center on general large language models often used for social or relational purposes (r/ChatGPT, r/ClaudeAI, r/OpenAI, r/GeminiAI, r/GoogleGeminiAI, r/grok, r/perplexity_ai). Across these subreddits, the gathered corpus through December 31, 2025 contained 24,695,601 entries (22,498,394 comments and 2,197,207 submissions). We then filtered this corpus to content containing the focal word *companion and companionship* between January 1, 2023 (not long after ChatGPT launch) and December 31, 2025, yielding 54,368 entries for the downstream analysis. This process is summarized in Figure 1.

We chose to include the broader term *companion* rather than restricting the corpus to *companionship* alone. Although *companion* includes more irrelevant content, it also substantially enlarges the pool of posts that may contain human–AI chat screenshots, which are comparatively rare but central to this study. We therefore treated this broader lexical filter as a retrieval strategy rather than a final relevance decision. Relevance was determined later through a companionship-topic coding method that paired companionship keywords from prior work (Banks et al., 2025) with a method for extending codes from limited validated data to a broader corpus (Banks et al., 2025). This approach allowed us to preserve image-bearing posts that a stricter companionship-only search would likely miss while reducing noise during downstream analysis.

From the 54,368 companion-filtered Reddit entries, we extracted any potential media URLs and downloaded associated images. This process yielded 4,411 images from 2,239 Reddit entries; of those, 1,547 images from 621 posts made by 473 users were identified as containing human–AI chat interactions and successfully transcribed after removing duplicated contents.

We leverage the multimodal LLM Qwen2.5-VL-72B-Instruct (Bai et al., 2025), served through vLLM (Kwon et al., 2023) to perform screenshot transcription. For each image, the model was prompted first to decide whether the image was a screenshot of a conversation between a human user and an AI chatbot, and then, if so, to transcribe only the visible chat bubbles in reading order. The prompt explicitly instructed the model to exclude non-conversation interface text such as app titles, timestamps, date dividers, menus, buttons, suggested replies, notifications, and other UI chrome. The final output included separate fields for the full transcript, user utterances, and bot utterances. A random $\approx$ 5% subset of transcripts ($n = 150$) was manually reviewed, indicating that the transcribed text was generally clean and free of interface noise. The main issue observed was occasional swapping of user and AI turns (3 out of 150); this problem was substantially reduced in a revised transcription effort by prompting the model to treat right-side chat bubbles as user messages when speaker assignment was uncertain (all 3 swapping errors in the sample were corrected).

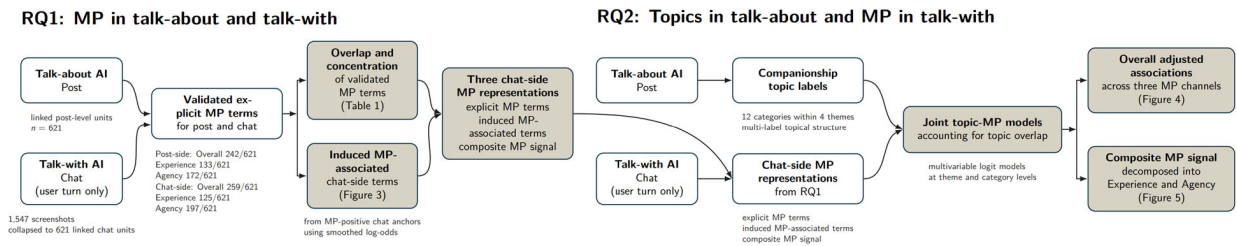

**Figure 2: Visual summary of analyses performed**

## 2.2 Analytical Approach

Our analytical strategy followed the conceptual distinction developed above between talk-about an AI companion and talk-with that AI. The Reddit post attached to a screenshot was treated as talk-about the experience with AI companion, whereas the transcribed user utterances in the screenshot were treated as talk-with the AI. This distinction allowed us to separate relatively explicit, topical companionship discussions and mind-attributing language in posts from more interactional, *in-situ* behavioral traces of mind perception in direct conversation.

*2.2.1 Identifying Post Topics in Talk-About.* First, we assigned companionship topics to Reddit posts using the 12-topic keyword set developed in past scholarship (Banks et al., 2025). We retained all topics rather than collapsing immediately to higher-order themes so that finer distinctions in companionship experience could be preserved for analysis. We evaluated that study's data (freely available through their online supplements) and found 75 screenshot posts overlapped with dataset overlapped. Because those posts had already been assigned topic-indicative terms in the prior study, we used them as a tuning set for adapting the prior keyword framework to this new corpus. We began with simple seed-keyword matching, which retrieved many relevant cases but also produced many false positives (weighted precision = .61, recall = .87, F1 = .70). We then expanded the keyword lists with additional high-coverage variants, which further increased recall (.96) but left precision modest (.58; weighted F1 = .71). To improve this balance, we applied a scoring approach that gives each topic a score based on its matched keywords, weighting rarer and more distinctive terms more heavily while adjusting for post length (Banks & Li, 2025).

Before scoring, post text was normalized by lowercasing, Unicode canonicalization, whitespace compression, and removal of punctuation not required for token boundaries. Seed matching used limited flexibility to improve recall without collapsing topic specificity: Prefix matching for starred stems, light morphological variation, and bounded in-order phrase windows for multiword expressions.

To reduce false negatives while preserving topic specificity, the overlap set was used to identify additional candidate seed terms for each topic. For a candidate term $t$ and topic $c$, let $U_t$ denote the set of support units containing $t$, $U_{tc}^+$ the support units coded with topic $c$ and containing $t$, and $U_c^+$ the overlap units coded with $c$. Candidate terms were evaluated with a precision-oriented proxy, PrecProxy$(t,c) = |U_{tc}^+|/|U_t|$, and a recall-oriented proxy, RecProxy$(t,c) = |U_{tc}^+|/|U_c^+|$. These were combined into

$$\text{Score}(t,c) = \text{PrecProxy}(t,c)\log(1 + |U_{tc}^+|)\big(1 + \text{IDF}(t)\big)\big(0.5 + \text{RecProxy}(t,c)\big),$$

where $\text{IDF}(t) = \log_2\big((|U| + 1)/(|U_t| + 1)\big)$. Only candidate terms meeting the minimum support requirement and a precision proxy of at least .80 were retained, after which the highest-scoring terms were appended to the topic seed lists.

Simple seed matching recovered many relevant cases, but it also tended to over-code long posts and topics represented by lexically common seeds. To improve topic assignment, we used a weighted scoring rule that down-weighted common topic vocabularies and adjusted for post length. Let $w_i$ be the word count of post $i$, $W = \sum_i w_i$ the corpus-wide post word count, and $T_c = \sum_i h_{ic}$ the corpus-wide hit total for topic $c$. Each topic received a corpus-level rarity scaler $q_c = \log_2(W/T_c)$ and each post–topic pair received a length-adjusted base score $b_{ic} = h_{ic}/w_i^\lambda$, where $\lambda \in [0,1]$ governs the strength of the length correction. The rarity-weighted topic score was $r_{ic} = b_{ic}\,q_c^\rho$, where $\rho \geq 0$ controls the contribution of rarity weighting. The notebook implementation allowed optional normalization of topic scores within a post, but the optimized final model selected the no-normalization option, so the downstream decision rule operated directly on $r_{ic}$.

Topics were first screened by minimum evidence requirements. Specifically, a topic was considered active only if it met the minimum number of matched seeds and distinct seed forms; in the optimized model, both minima were one, so any topic with at least one matched seed remained eligible. Among active topics, codes were assigned relative to the within-post score distribution. Let $\sigma_i$ be the standard deviation of positive topic scores within post $i$. The post-specific selection threshold was

$$\tau_i = \max\left(\eta, \ \max_{c \in \mathcal{C}_i^+} r_{ic} - \alpha \sigma_i\right),$$

where $\mathcal{C}_i^+$ denotes the set of active topics in post $i$, $\eta$ is a minimum absolute score floor, and $\alpha$ controls how far below the within-post maximum a topic may fall while still being retained. Topics were retained when their scores met or exceeded $\tau_i$, with a hard cap of at most $L_{\max} = 12$ codes per post.

The score parameters were tuned against the 75-post overlap set using 500 Bayesian optimization trials. The optimized parameter vector was $\theta = \{\rho, \lambda, \alpha, \eta, L_{\max}\}$, subject to the constraints defined in the notebook search space. For a candidate parameterization $\theta$, the predicted multilabel assignment matrix $\hat{\mathbf{Y}}(\theta)$ was compared with the accepted-topic assignment matrix $\mathbf{Y}$. The optimization target favored recall while still rewarding precision for broader coverage,

$$\mathcal{O}(\theta) = 0.3\,\mathrm{Precision}_w(\theta) + 0.7\,\mathrm{Recall}_w(\theta),$$

where the subscript $w$ denotes the weighted-average multi-topic metric over the 12 topics. Weighted F1 was reported for interpretive completeness but was not itself optimized. Relative to seed matching alone, the final optimized model improved performance to weighted precision = .81, recall = .95, and F1 = .87. We then applied these optimized settings to the full screenshot-linked post corpus to assign companionship topics, which served both as the topical measurement layer for companionship and as a relevance filter for the downstream post–chat linkage.

### 2.2.2 Identifying MP Language in Talk-With and Talk-About.

Second, we operationalized mind perception by leveraging the Mind Perception Dictionary (MPD; Schweitzer & Waytz, 2021), focusing on Experience and Agency lexica while treating Mind Overall (inclusive of experience and agency) as a composite signal. We first applied the dictionary to Reddit post text to derive post-side MP indicators in talk about AI. These candidate hits were then contextually validated case-by-case by GPT 5.4 to determine whether the matched terms actually functioned as indicators of experience and/or agency in context. This helps to ensure the MPD terms actually semantically indicated MP, for instance distinguishing between the lemma *passion\** manifesting as "*passion*ate response" (retained) and "*passion* fruit" (excluded).

The model was instructed to evaluate each matched term in its local sentence or turn context and retain it only when the wording attributed a mental capacity or subjective state to the AI itself (and not some other meaning). MP terms were retained only when the wording attributed experiential or agentic mental states to the AI. Hits were rejected when the matched term referred to the user, another human, a hypothetical person, generic people, interface text, routine conversational fillers, or metaphorical or idiomatic language alone; ambiguous cases defaulted to rejection. Overall MP was identified as present whenever either Experience or Agency was valid in context. This procedure yielded validated post-side and chat-side terms for overall mind perception, experience, and agency. In manual audits of 10% samples from the post-side and chat-side candidates containing MP terms, reviewer noted that there are missed valid terms occurred in 7 of 28 post-side cases and 10 of 28 chat-side cases (17/56 overall, 30.4%), suggesting that the model was reliable at excluding clearly invalid matches but somewhat selective, sometimes dropping valid indicators; this is recognized as a limitation.

### 2.2.3 Identifying Topical and MP Language Relationships.

To address the posed research questions, we systematically analyzed the relationships between the pertinent linguistic indicators across both corpora (Figure 2).

For **RQ1**, we evaluated the relationship between MP indicators in talk-with and MP indicators in talk-about the AI companion. For talk-with language, we analyzed only the user side of the transcribed chats so that chat-based MP signals reflected the human's language rather than the AI companion's replies (which are generally designed to reflect anthropomorphic mental states). Building on the validated post-side and chat-side MP terms described above, we first used the validated chat-side MP terms as a baseline check on whether posts that with MP terms also tended to be linked to chats with MP terms. We then summarized explicit dictionary-based MP prevalence at two scopes: The set of posts with usable linked chat text (621 posts) and the usable chat units themselves (621 user-side chat units after merging all screenshots from the same Reddit post into a single linked chat record). Posts with multiple screenshots typically represented longer stretches of the same conversation rather than distinct independent interactions, so this merge preserved a more complete chat record while ensuring that each Reddit post contributed only one case to the analysis.

To characterize how explicit MP terms overlapped across contexts, the analysis used set-based and concentration-based summaries. Corpus-level lexical overlap was quantified with Jaccard similarity over the sets of unique validated terms observed in post-side and chat-side units for each dimension: if $A_d^{\text{post}}$ and $A_d^{\text{chat}}$ denote the unique validated terms for dimension $d$ in posts and chats, respectively, then

$$J_d = \frac{\left|A_d^{\text{post}} \cap A_d^{\text{chat}}\right|}{\left|A_d^{\text{post}} \cup A_d^{\text{chat}}\right|}.$$

Term concentration within a context was summarized with the Herfindahl–Hirschman Index (HHI), $\text{HHI} = \sum_t p_t^2$, where $p_t$ is the proportional share of validated hits contributed by term $t$, alongside the Top-5 share, defined as the proportion of all validated hits accounted for by the five most frequent terms. Higher HHI values indicate concentration in a smaller number of recurrent terms.

The validated MPD terms provided a baseline for explicit MP language, but they were unlikely to capture all ways that MP may be expressed in direct interaction with an AI. In chat, mindedness may also be conveyed through situated and relational phrasing, including direct address, epistemic constructions, and recurrent interactional expressions that do not appear as standalone dictionary terms. Unlike the post-side topic-matching pipeline, the discovery stage permitted both unigrams and adjacent bigrams. Candidate collocation bigrams were screened with the Dunning log-likelihood ratio statistic, which compares the independence model with the collocation model for a $2 \times 2$ contingency table; in compact form, $\text{LLR} = 2\left[\log L(H_1) - \log L(H_0)\right]$, where larger values indicate stronger evidence that the two words co-occur more often than expected under independence. Bigrams were retained when they met the post thresholds of count $\geq 3$, support users $\geq 2$, and LLR $\geq 6.63$ (corresponds roughly to $p < .01$), while also containing interpretable content rather than only generic platform or referent terms.

To identify such indicators, we needed a method that would compare MP-positive and MP-negative chat units and recover terms that are disproportionately characteristic of the positive class rather than merely common overall. We therefore used smoothed log-odds with an informative Dirichlet prior (Monroe et al., 2008). This approach was selected because chat units may be short and lexically sparse, making raw frequency contrasts unstable. Relative to unsmoothed log-odds, this regularizes low-count terms by shrinking them toward background corpus frequencies. Relative to Pointwise Mutual Information (PMI), which tends to overvalue rare co-occurrences and does not directly estimate class-distinctiveness, the selected approach tends to yield a direction-specific and variance-adjusted measure of lexical distinctiveness. This makes it well suited for identifying interpretable indicators of mind perception in conversational text.

Specifically, discovery was performed separately for Experience and Agency, using the validated chat-side terms as anchors. For dimension $d$, let $x_{td}^+$ and $x_{td}^-$ be the counts of token $t$ in the MP-positive and MP-negative training subsets, respectively. With symmetric additive prior $\alpha = 0.01$, define $\tilde{x}_{td}^+ = x_{td}^+ + \alpha$, $\tilde{x}_{td}^- = x_{td}^- + \alpha$, $\tilde{N}_d^+ = \sum_t \tilde{x}_{td}^+$, and $\tilde{N}_d^- = \sum_t \tilde{x}_{td}^-$. The smoothed log-odds contrast for token $t$ was

$$\delta_{td} = \log\left(\frac{\tilde{x}_{td}^+}{\tilde{N}_d^+ - \tilde{x}_{td}^+}\right) - \log\left(\frac{\tilde{x}_{td}^-}{\tilde{N}_d^- - \tilde{x}_{td}^-}\right),$$

$$z_{td} = \frac{\delta_{td}}{\sqrt{1/\tilde{x}_{td}^+ + 1/\tilde{x}_{td}^-}}.$$

This regularizes low-count terms toward background corpus frequencies and yields a direction-specific measure of lexical distinctiveness that is more stable than raw frequency contrasts in short chat documents.

Candidate tokens were required to satisfy a positive-direction criterion ($\delta_{td} > 0$), the post significance threshold ($z_{td} > 1.96$), and a minimum positive-support threshold of at least two support users. Support was counted at the user level when author metadata were available; when author information was missing, support defaulted to one support unit per post to avoid inflating recurrence from duplicated rows.

To reduce dependence on a single split of the data, token selection was filtered by repeated user-level subsampling. Let $B = 80$ denote the number of stability iterations. At each iteration $b$, 80% of unique support users were sampled without replacement, token statistics were recomputed, and a token was marked selected if it again met the positive-direction, $z$, and support thresholds. The stability rate for token $t$ was $\text{Stab}(t) = \frac{1}{B}\sum_{b=1}^{B} \mathbb{I}\left(t \text{ selected in iteration } b\right)$, and tokens were retained only when $\text{Stab}(t) \geq .60$.

A held-out replication check further required positive distinctiveness in a user-grouped holdout split. The discovery pipeline used a GroupShuffleSplit with 30% of support users assigned to holdout whenever multiple users were available. Let $\delta_{td}^{\text{hold}}$ denote the held-out log-odds contrast and $s_{td,+}^{\text{hold}}$ the held-out positive support for token $t$. Replication required $\delta_{td}^{\text{hold}} > 0$ and $s_{td,+}^{\text{hold}} \geq 2$. The retained indicator pool for each dimension therefore consisted of tokens that were positively distinctive, sufficiently recurrent, stable across user-level resampling, and directionally replicated in holdout.

For each dimension $d$, the retained token set $\mathcal{S}_d$ was converted into a chat-level score. Let $m_{it} = \mathbb{I}(t \in u_i)$ indicate whether token $t$ appears at least once in user-side chat unit $i$, and let $\ell_i$ denote the word count of that chat unit. Tokens were weighted by the nonnegative portion of their standardized log-odds score, $\omega_{td} = \max(z_{td}, 0)$, and the latent chat-side MP score for unit $i$ and dimension $d$ was

$$g_{id} = \frac{\sum_{t \in \mathcal{S}_d} m_{it}\ \omega_{td}}{\ell_i^{\lambda_{\text{MP}}}},$$

with $\lambda_{\text{MP}} = .50$ to dampen document-length effects. The decision threshold for the latent score was prevalence-matched on the training data. If $\pi_d$ denotes the explicit positive rate for dimension $d$ in training, then the threshold was the $(1 - \pi_d)$ empirical quantile of the training score distribution, $\kappa_d = Q_{1-\pi_d}(\{g_{id} : i \in \mathcal{I}_{\text{train}}\})$. The latent indicator for unit $i$ was $\hat{Y}_{id}^{\text{latent}} = \mathbb{I}(g_{id} \geq \kappa_d)$, and the composite chat-side MP signal was defined as $\hat{Y}_{id}^{\text{comp}} = \max(Y_{id}, \hat{Y}_{id}^{\text{latent}})$. This yielded three post-facing chat-side MP representations for downstream analysis: validated explicit MP terms, induced MP-associated terms, and their composite.

The **RQ1** analyses yielded three related representations of chat-side MP: Validated explicit MP terms, induced MP-associated terms, and a composite signal combining the two. These representations allowed **RQ2** to test whether companionship topics related not only to chat-side MP overall, but also to different forms of its expression in talk-with AI.

This design follows directly from the **RQ1** finding that chat-side MP is not exhausted by validated MPD terms alone.

To address **RQ2**, we linked the companionship-topic codes derived from Reddit posts to the associated user-side chat analysis units and examined whether post-side companionship topics were related to chat-side MP in the linked conversations. The topical measure consisted of the 12 companionship topics and their corresponding 4 higher-order topical themes. Because these topics overlap rather than function as mutually exclusive classes, we modeled them jointly rather than treating them as independent.

We first summarized the prevalence of themes and topics within the 621 linked post–chat units to characterize the topical composition of the analytic corpus. Confidence intervals for these proportions used the Wilson interval, which provides better small-sample behavior than the simple Wald interval. For an observed proportion $\hat{p} = x/n$ and standard-normal critical value $z_{.975}$, the 95% Wilson interval was computed as

$$\frac{\hat{p} + z_{.975}^2/(2n)}{1 + z_{.975}^2/n} \pm \frac{z_{.975}}{1 + z_{.975}^2/n} \sqrt{\frac{\hat{p}(1-\hat{p})}{n} + \frac{z_{.975}^2}{4n^2}},$$

where the quantity being estimated is the prevalence of each companionship theme or topics in the linked corpus.

We then estimated adjusted associations between companionship topics and chat-side MP using logistic regression. Separate models were fit at the theme and topics levels, and separately for each chat-side MP representation established in **RQ1**: validated explicit MP terms, induced MP-associated terms retained on the basis of positive class distinctiveness, z > 1.96, the minimum support, selection stability across resampling, and positive held-out replication, and a composite signal combining the two. Let $X_{ic}$ denote a binary indicator that topic $c$ is present in post $i$. Separate logistic regressions were fit for each chat-side MP representation—explicit, induced, and composite:

$$\text{logitPr}\left(M_i^{(k)} = 1\right) = \beta_0^{(k)} + \sum_{c=1}^{C} \beta_c^{(k)} X_{ic},$$

where $M_i^{(k)}$ is the chat-side MP outcome for unit $i$ and $C$ is either the number of themes or the number of topics, depending on model level. Because all topic indicators entered the model simultaneously, coefficient $\beta_c^{(k)}$ estimates the association between topic $c$ and chat-side MP after adjusting for co-occurrence with the remaining topics. We report coefficient estimates and 95% confidence intervals based on heteroskedasticity-robust standard errors. We then decomposed the composite chat-side MP signal into its Experience and Agency dimensions to test whether topical relationships differed by MP dimension rather than only by overall presence. For dimension $d \in \{E, A\}$, the decomposition model was

$$\text{logitPr}\left(M_{id}^{(\text{comp})} = 1\right) = \gamma_0^{(d)} + \sum_{c=1}^{C} \gamma_c^{(d)} X_{ic}.$$

Together, these analyses tested whether companionship topics in talk-about AI were associated with linguistic indicators of MP in talk-with AI, and whether those associations differed by the form in which MP was expressed.

## 3  Results

The analytic corpus for these results consists of 621 linked post–chat units; each unit included one post-side "talk-about" record and one merged chat-side "talk-with" record per Reddit post. At a high level, post-side language was more lexically varied and descriptive, whereas chat-side language was shorter, more interactional, and more concentrated around a smaller set of recurrent terms. The corpus was subjected to analyses as described above, forwarding results as described below. Throughout these results, "talk-about" AI companions corresponds with post-side data, while "talk-with" AI companions corresponds with chat-side data.

### 3.1  Linguistic Indicators of MP in Chat with AI Companions (RQ1)

Using rote terms from the Mind Perception Dictionary (Schweitzer & Waytz, 2021) after excluding generic, metaphorical, off-topic, or otherwise non-attributive terms, overall MP appeared in 40.5% of screenshot-inclusive posts (242/621; 68.1% before validation), while overall MP appeared in 41.7% of linked chat units (259/621; 65.5% before validation). The same pattern held across dimensions: Experience MP appeared in 21.4% of posts (133/621; 53.8% before validation) and in 20.1% of chats (125/621; 45.6% before validation); Agency MP appeared in 27.7% of posts (172/621; 57.3% before validation) and in 31.7% of chats (197/621; 53.1% before validation). Overall, post-side and chat-side MP-term rates were broadly comparable among screenshot-inclusive posts, with chat-side Agency MP terms being somewhat more prevalent.

Table 1 shows specific MP terms. In posts, the most common validated dictionary hits were higher-order Experience MP terms such as *feel*, *emotion*, and *experience*, and Agency MP terms like *think* and *understand*. Chats, on the other hand, were more concentrated around a smaller set of high-frequency terms, especially *feel* for Experience MP and *think* for Agency MP. This difference suggests that rote MP terms in talk-about AI drew from a wider descriptive vocabulary while talk-with AI relied on a narrower set of recurrent mindedness terms.

**Table 1. Validated Mind Perception Dictionary term hits in talk-about (Reddit posts) and talk-with (chat-side conversations with AI), shown separately for Experience and Agency dimensions of MP.**

| MP Dimension | Post-Side MP term hits (count) | Chat-Side AI MP term hits (count) | All Overlapping MP terms in Alphabetical order |
|---|---|---|---|
| **Experience**<br><br>MPD total unique terms:<br>223 | Unique Term Coverage: 114/223 = 51.1%<br>Top-5 Share: 48.2%<br>HHI: 0.068 | Unique Term Coverage: 36/223 = 16.1%<br>Top-5 Share: 72.2%<br>HHI: 0.179 | Unique Overlapping Terms: 32<br>Jaccard Overlap (unique MP terms): 0.271 |
| | feel* (146); emotion* (113); experience* (75); care* (31); hope* (27); desire* (20); enjoy* (18); fear* (16); happy (16); sad* (15); patien* (14); calm (12); empath* (12); fascinat* (12); hurt* (12); concern* (11); content (11); lonely (11); affection (9); surprise* (9); honor (8); interested (8); crav* (7); devot* (7); tired (7); inspired (6); mood (6); afraid (5); eager* (5); perceive* (5); solace (5); aspir* (4); confident (4); excited (4); glad (4); grief (4); joy* (4); pain (4); sorry (4); strong (4); thrilled (4); worried (4); admir* (3); caring (3); clever (3); comfortable (3); motivation (3); passion (3); scared (3); shame (3); angry (2); annoyed (2); anxiety (2); awe (2); compassion (2); daring (2); digni* (2); discomfort (2); hunger (2); inspiration (2); lively (2); miser* (2); nervous (2); perception (2); pride* (2); proud (2); rage (2); relie* (2); serene (2); suffer* (2); upset (2); wise (2); absorbed (1); advers* (1); aggressive (1); alarmed (1); amuse* (1); anger (1); appetite (1); apprehensive (1); astounded (1); attracted (1); avid (1); blush* (1); bold (1); brave (1); cheerful (1); contentious (1); cynical (1); dedicated (1); delight* (1); disdain* (1); disgust* (1); distress* (1); earnest (1); enthusiastic (1); frustrated (1); horrified (1); hostile (1); jovial (1); keen (1); mad (1); modest (1); obsess* (1); pleasure (1); poised (1); regret* (1); relaxed (1); satisfaction (1); shocked (1); shy (1); sorrow* (1); temper (1); wound* (1) | feel* (60); hope* (18); enjoy* (15); care* (13); happy (8); emotion* (4); strong (3); desire* (2); eager* (2); experience* (2); frustrated (2); mad (2); patien* (2); proud (2); surprise* (2); admir* (1); blush* (1); comfortable (1); concern* (1); crav* (1); delight* (1); empath* (1); excited (1); fascinat* (1); fear* (1); frightened (1); glad (1); hurt* (1); inspiration (1); lonely (1); miser* (1); sad* (1); satisfied (1); scared (1); timid (1); trembling (1) | admir* (post 3; chat 1); blush* (post 1; chat 1); care* (post 31; chat 13); comfortable (post 3; chat 1); concern* (post 11; chat 1); crav* (post 7; chat 1); delight* (post 1; chat 1); desire* (post 20; chat 2); eager* (post 5; chat 2); emotion* (post 113; chat 4); empath* (post 12; chat 1); enjoy* (post 18; chat 15); excited (post 4; chat 1); experience* (post 75; chat 2); fascinat* (post 12; chat 1); fear* (post 16; chat 1); feel* (post 146; chat 60); frustrated (post 1; chat 2); glad (post 4; chat 1); happy (post 16; chat 8); hope* (post 27; chat 18); hurt* (post 12; chat 1); inspiration (post 2; chat 1); lonely (post 11; chat 1); mad (post 1; chat 2); miser* (post 2; chat 1); patien* (post 14; chat 2); proud (post 2; chat 2); sad* (post 15; chat 1); scared (post 3; chat 1); strong (post 4; chat 3); surprise* (post 9; chat 2) |
| **Agency**<br><br>MPD total unique terms:<br>104 | Unique Term Coverage: 72/104 = 69.2%<br>Top-5 Share: 27.7%<br>HHI: 0.031 | Unique Term Coverage: 37/104 = 35.6%<br>Top-5 Share: 65.2%<br>HHI: 0.174 | Unique Overlapping Terms: 36<br>Jaccard Overlap (unique MP terms): 0.493 |
| | think* (76); understand* (75); love* (70); imagin* (44); intelligen* (42); thought* (42); believe* (38); inten* (36); memory (35); plan* (35); aware* (33); decide* (31); communicat* (28); conscious* (25); prefer* (25); remembers (25); reason* (23); recogni* (22); goal* (21); control (20); forget* (20); purpose (19); realize* (19); mental* (18); accept* (17); predict* (17); value (17); brain* (16); likes (15); mind* (15); agen* (13); judge* (13); ethical (12); remembered (11); evil (10); perspective (10); forgot* (9); opinion (9); recall* (9); appreciate (8); intellect* (7); moral* (7); focused (6); responsible (6); impressed (5); prepare* (5); abus* (4); envision* (4); attitude* (3); competen* (3); fair (3); visualiz* (3); adore* (2); calculate* (2); conclude* (2); deliberate (2); determined (2); infer* (2); organize* (2); responsive (2); attentive (1); conceive* (1); discretion (1); dislikes (1); formulate* (1); liking (1); memorize* (1); noble (1); rational (1); restrain* (1); unethical (1); unfair (1) | think* (109); love* (21); understand* (21); mind* (20); thought* (13); believe* (12); realize* (9); plan* (7); memory (6); opinion (5); prefer* (5); aware* (4); brain* (4); decide* (4); imagin* (4); intelligen* (4); accept* (3); communicat* (3); conscious* (3); perspective (3); value (3); appreciate (2); forget* (2); infer* (2); control (1); evil (1); foresee* (1); impressed (1); inten* (1); mental* (1); predict* (1); prepare* (1); purpose (1); reason* (1); recall* (1); recogni* (1); remembered (1) | accept* (post 17; chat 3); appreciate (post 8; chat 2); aware* (post 33; chat 4); believe* (post 38; chat 12); brain* (post 16; chat 4); communicat* (post 28; chat 3); conscious* (post 25; chat 3); control (post 20; chat 1); decide* (post 31; chat 4); evil (post 10; chat 1); forget* (post 20; chat 2); imagin* (post 44; chat 4); impressed (post 5; chat 1); infer* (post 2; chat 2); intelligen* (post 42; chat 4); inten* (post 36; chat 1); love* (post 70; chat 21); memory (post 35; chat 6); mental* (post 18; chat 1); mind* (post 15; chat 20); opinion (post 9; chat 5); perspective (post 10; chat 3); plan* (post 35; chat 7); predict* (post 17; chat 1); prefer* (post 25; chat 5); prepare* (post 5; chat 1); purpose (post 19; chat 1); realize* (post 19; chat 9); reason* (post 23; chat 1); recall* (post 9; chat 1); recogni* (post 22; chat 1); remembered (post 11; chat 1); think* (post 76; chat 109); thought* (post 42; chat 13); understand* (post 75; chat 3); value (post 17; chat 3) |

Counts indicate the number of validated positive analysis units in each context containing the term at least once. Rankings are based on within-context term frequency among validated positive cases. Top-5 share is the percentage of all validated term hits in a context accounted for by the five most frequent validated terms. Jaccard overlap is computed over unique validated MP terms. HHI is computed over the full distribution of validated MP term hits in a context by summing the squared proportional shares of each term; higher values indicate greater concentration.

Even when MP appeared on both sides, correspondence was substantial but incomplete rather than absent. Table 1 shows considerable corpus-level lexical overlap between validated post-side and chat-side MP terms: For Experience MP, 32 terms overlapped between 114 post-side and 36 chat-side unique terms (Jaccard = 0.271); For Agency MP, 36 terms overlapped between 72 post-side and 37 chat-side unique terms (Jaccard = 0.493). This shared core lexicon included Experience-related terms such as *feel, hope,* and *care* and Agency-related terms such as *think, understand, love,* and *memory.* At the same time, the two contexts were not lexically identical. Post-side MP drew on a broader descriptive vocabulary, whereas chat-side MP was more concentrated around a smaller set of recurrent terms. For Experience, the five most frequent validated terms accounted for 48.2% of all post-side term hits but 72.2% of chat-side term hits (HHI = 0.068 vs. 0.179). For Agency, the corresponding shares were 27.7% for posts and 65.2% for chats (HHI = 0.031 vs. 0.174). Taken together, this pattern suggests continuity in the core lexicon of mind perception across contexts, alongside a narrowing and concentration of that lexicon in direct interaction with AI.

Next, using smoothed log-odds with an informative Dirichlet prior (Monroe et al. 2008), we identified terms that were especially distinctive of chat-side MP. As shown in Figure 3, these indicators included both theory-aligned MP terms and more interactional expressions. In the Experience plot, prominent indicators such as *feel, care, surprise,* and *hope* align closely with the Experience dimension of the MP Dictionary. In the Agency plot, terms such as *think, memory, believe* and *understand* are also consistent with the dictionary's Agency dimension, which includes cognition-, memory-, and reasoning-related language. At the same time, several distinctive indicators were embedded in conversational or relational phrasing rather than appearing only as abstract dictionary-style words, including forms such as *with you, you say, to hear,* and *our.* This pattern suggests that explicit mind perception in talk-with AI was expressed not only in overt declarations that the AI has a mind but also, and often, in language suggestive of interactional stance, mutual reference, and direct engagement.

## 3.2 Links Between Companionship Topics and MP Indicators in AI Companion Chat (RQ2)

To explore how in-chat MP indicators may be useful in better understanding companionship experiences, we identified relationships between the *subject matter* of talk-about AI companions (i.e., facets of companionship indicated by known keywords, as reported in Banks et al., 2025) and the above-validated MP indicators in talk-with those companions. The subject matter is organized by topical themes and their more granular topics.

Within the 621 linked post–chat units, shown in Table 2 and Figure 4, Socioemotionality was the most prevalent theme (68.3%), followed by User Control (43.8%), Imaginaries (28.0%), and Limitations (22.1%). At the theme level, Realism (53.6%) and Bonding (47.5%) were the most common, with Playfulness (26.6%), Customization (20.9%), and Existential/Philosophical Issues (17.1%) also appearing frequently. However, high prevalence did not necessarily translate into strong independent relationships with chat-side MP. After adjustment for topic co-occurrence, the largest topics such as Bonding and Realism remained close to zero across the explicit, induced, and composite channels, with confidence intervals spanning zero for Bonding, Realism, and Sex(uality). A similar pattern held for User Control, which was generally flat to slightly negative across channels, with

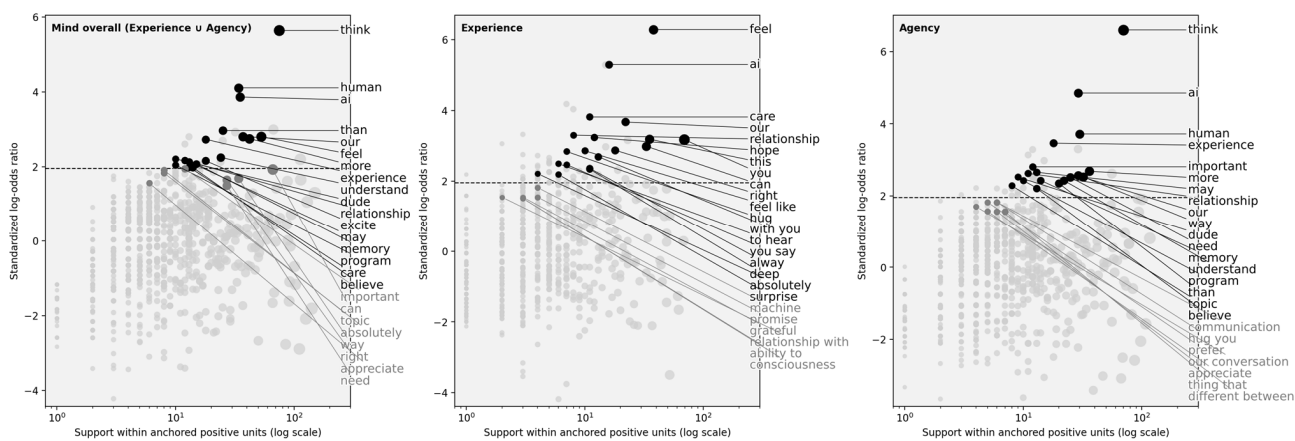

**Figure 3: Distinctive chat-side MP lexical indicators for Mind Overall, Experience, and Agency MP, estimated using smoothed log-odds with an informative Dirichlet prior.**

The y-axis shows the standardized log-odds ratio (Monroe et al., 2008), where larger positive values indicate that a term is more distinctive of MP-positive chat units for that dimension. The x-axis shows support, defined as the number of anchored positive units containing the term. The dashed horizontal line marks $z = 1.96$. Labeled terms are the top 25 indicators based on standardized log-odds ratio and support shown in the panel.

**Table 2. Theme- and topic-level prevalence and adjusted associations between companionship topics and chat-side MP.**

| Theme / Topic | n | Prev (%) | Explicit MP terms | | Induced MP-associated terms | | Composite MP terms | |
|---|---|---|---|---|---|---|---|---|
| | | | % with 95% CI | Adj. log-odds [95% CI] | % with 95% CI | Adj. log-odds [95% CI] | % with 95% CI | Adj. log-odds [95% CI] |
| **Socioemotionality** | 424 | 68.3 | 38.4 [33.9, 43.2] | -0.12 [-0.64, 0.40] | 37.7 [32.2, 42.4] | 0.36 [-0.12, 0.84] | 50.9 [46.2, 55.7] | -0.04 [-0.52, 0.43] |
| Bonding | 295 | 47.5 | 39.3 [33.9, 45.0] | 0.02 [-0.45, 0.49] | 37.6 [32.3, 43.3] | 0.09 [-0.33, 0.51] | 51.2 [45.5, 56.8] | -0.09 [-0.52, 0.33] |
| Realism | 333 | 53.6 | 39.3 [34.2, 44.7] | 0.03 [-0.44, 0.51] | 37.2 [32.2, 42.5] | -0.00 [-0.44, 0.43] | 52.0 [46.6, 57.3] | 0.05 [-0.38, 0.47] |
| Sex(uality) | 101 | 16.3 | 37.6 [28.8, 47.4] | -0.05 [-0.59, 0.49] | 36.6 [27.9, 46.4] | 0.02 [-0.50, 0.53] | 50.5 [40.9, 60.0] | -0.06 [-0.55, 0.43] |
| **User Control** | 272 | 43.8 | 39.7 [34.1, 46.5] | 0.07 [-0.37, 0.51] | 35.3 [29.9, 41.1] | -0.24 [-0.64, 0.17] | 50.7 [44.8, 56.6] | -0.10 [-0.50, 0.30] |
| Customization | 130 | 20.9 | 33.1 [25.6, 41.5] | -0.36 [-0.90, 0.17] | 30.8 [23.5, 39.2] | -0.40 [-0.88, 0.08] | 47.7 [39.3, 56.2] | -0.21 [-0.70, 0.28] |
| Playfulness | 165 | 26.6 | 41.2 [34.0, 48.8] | 0.01 [-0.47, 0.49] | 37.0 [30.0, 44.6] | -0.02 [-0.45, 0.40] | 51.5 [43.9, 59.0] | -0.15 [-0.59, 0.29] |
| Boundary negotiation | 82 | 13.2 | 41.5 [31.4, 52.3] | 0.37 [-0.26, 0.99] | 40.2 [30.3, 51.1] | 0.14 [-0.43, 0.70] | 54.9 [44.1, 65.2] | 0.27 [-0.30, 0.84] |
| **Limitations** | 140 | 22.5 | 45.0 [37.0, 53.3] | 0.65 [0.18, 1.12] | 45.7 [37.7, 54.0] | 0.66 [0.22, 1.09] | 57.9 [49.6, 65.7] | 0.50 [0.06, 0.95] |
| Inauthenticity | 26 | 4.2 | 65.4 [46.2, 80.6] | 1.31 [0.40, 2.23] | 57.7 [38.9, 74.5] | 0.72 [-0.12, 1.56] | 73.1 [53.9, 86.3] | 0.84 [-0.03, 1.71] |
| Transactionality | 82 | 13.2 | 43.9 [33.7, 54.7] | 0.56 [-0.02, 1.15] | 43.9 [33.7, 54.7] | 0.57 [0.05, 1.08] | 58.5 [47.7, 68.6] | 0.57 [0.03, 1.11] |
| Ethicality | 52 | 8.4 | 36.5 [24.8, 50.1] | -0.09 [-0.81, 0.63] | 40.4 [28.2, 53.9] | 0.15 [-0.47, 0.76] | 46.2 [33.3, 59.5] | -0.33 [-0.97, 0.32] |
| **Imaginaries** | 174 | 28.0 | 40.2 [33.2, 47.7] | -0.02 [-0.48, 0.44] | 41.4 [34.3, 48.8] | 0.26 [-0.16, 0.68] | 55.2 [47.8, 62.4] | 0.20 [-0.23, 0.63] |
| Social Isolation | 46 | 7.4 | 41.3 [28.3, 55.7] | 0.00 [-0.71, 0.71] | 47.8 [34.1, 61.9] | 0.42 [-0.22, 1.06] | 56.5 [42.2, 69.8] | 0.08 [-0.59, 0.74] |
| Speculation | 66 | 10.6 | 37.9 [27.1, 49.9] | -0.03 [-0.68, 0.62] | 34.8 [24.5, 46.9] | -0.14 [-0.72, 0.43] | 50.0 [38.3, 61.7] | -0.08 [-0.67, 0.51] |
| Exist./Philos. | 106 | 17.1 | 39.6 [30.8, 49.1] | -0.00 [-0.57, 0.56] | 46.2 [37.0, 55.7] | 0.55 [0.07, 1.03] | 59.4 [49.9, 68.3] | 0.50 [-0.01, 1.01] |

Companionship topics were assigned from post text, and MP was measured from the linked user-side chat text in the 621 linked post–chat units. For each theme and topic, the table reports the number of labeled units, observed prevalence with 95% confidence intervals, and adjusted log-odds estimates with 95% confidence intervals for three chat-side MP representations: validated explicit MP terms, induced MP-associated terms, and the composite of the two. Adjusted estimates come from multivariable binomial generalized linear models in which all topic indicators were entered simultaneously, so each coefficient represents the association of that topic with chat-side MP while accounting for overlap with the remaining topics. Theme rows summarize higher-order topical groupings, and indented rows report the corresponding topic-level estimates. Because companionship topics are multi-label, counts and percentages do not sum to 100 across rows.

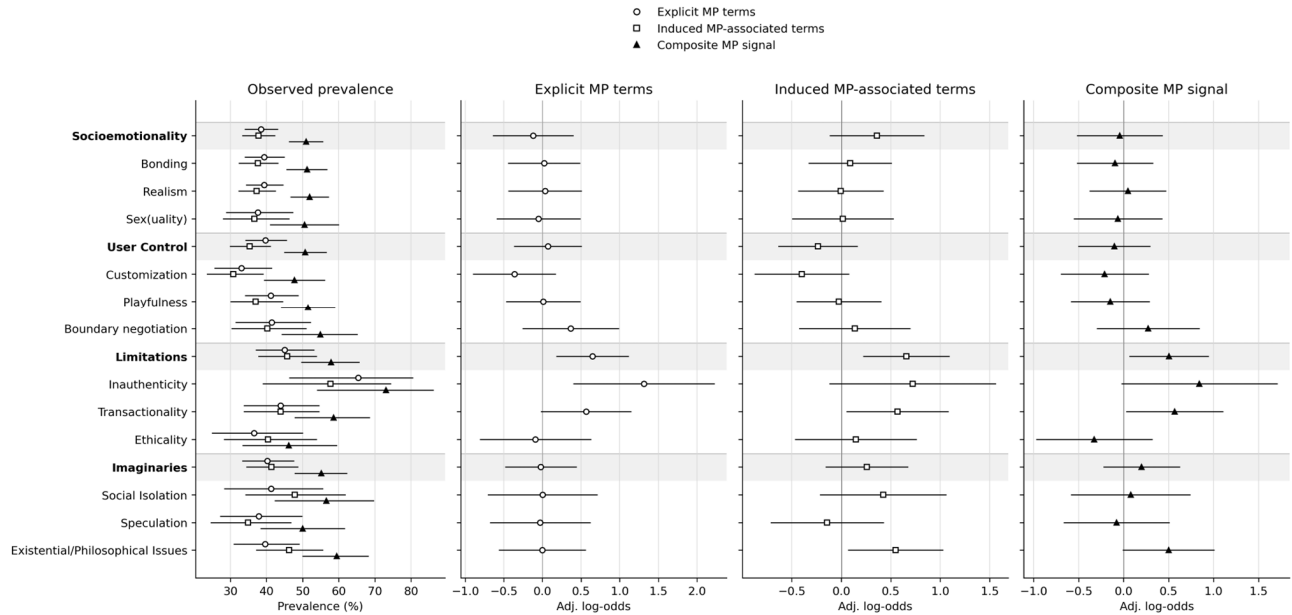

**Figure 4: Observed prevalence and adjusted overall MP associations for companionship themes and topics in the linked post/chat units**

Theme-level rows and indented topic-level rows show companionship topics assigned from Reddit post text, whereas MP was measured from the linked user-side chat text. The left panel reports observed prevalence (% of the 621 linked post–chat units) with 95% confidence intervals. The three right panels report adjusted log-odds estimates and 95% confidence intervals from multivariable binomial generalized linear models fit separately for validated explicit MP terms, induced MP-associated terms, and the composite MP terms. Topic indicators were entered simultaneously within each model, so each estimate reflects the association of a given topic with chat-side MP while accounting for co-occurrence with the other topics. Positive coefficients indicate greater likelihood of chat-side MP for units carrying that topic, net of overlap with the remaining topics.

the weakest estimates appearing for Customization. Taken together, these results suggest that the most common companionship frames were not, by themselves, the strongest independent correlates of chat-side MP.

The clearest positive adjusted associations instead appeared in Limitations. At the theme level, Limitations showed a positive association with explicit MP terms (adjusted log-odds = 0.65, 95% CI [0.18, 1.12]) and with induced MP-associated terms (0.66, [0.22, 1.09]), while the composite terms remained positive but less certain (0.50, [0.06, 0.95]). This pattern was driven most clearly by Inauthenticity, which had the largest explicit-MP estimate in the table (1.31, [0.40, 2.23]) and also elevated composite prevalence (73.1%, 95% CI [53.9, 86.3]). Transactionality showed a smaller but still positive explicit association (0.56, [-0.02, 1.15]), whereas Ethicality remained comparatively weak. Thus, limitation-oriented talk appeared more closely tied to chat-side MP than the more prevalent socioemotional topics, especially when MP was measured conservatively through validated explicit MP terms.

Imaginaries showed a more modest but suggestive pattern. As a theme, Imaginaries was near zero for explicit MP but trended positive for induced MP-associated terms (0.26, [-0.16, 0.68]) and for the composite terms (0.20, [-0.23, 0.63]). At the theme level, this pattern was driven mainly by Existential/Philosophical Issues, which showed the strongest induced-term association within the theme (0.55, [0.07, 1.03]) and a similarly positive but less certain composite estimate (0.50, [-0.01, 1.01]). By contrast, Speculation remained near zero or slightly negative across channels, while Social Isolation was somewhat more positive but imprecisely estimated. This suggests that not all imaginative companionship talk related to MP in the same way; rather, existential or philosophical framings were more closely linked to chat-side MP than speculative ones.

Figure 5 further shows that, when the composite chat-side MP terms was decomposed into Experience and Agency, most topic-level differences remained modest and uncertainty remained substantial. Even so, the decomposition suggested that the clearest positive patterns in Limitations, especially Inauthenticity, were visible across both dimensions, whereas Existential/Philosophical Issues appeared somewhat more agentic than experiential. In contrast, the dominant socioemotional topics again remained comparatively flat. Overall, the results indicate that chat-side MP was broadly distributed across companionship topics rather than tightly concentrated in a single topical niche, but that limitation-oriented talk—and to a lesser extent existential/philosophical talk—showed the positive relationships with linguistic indicators of MP in talk-with AI.

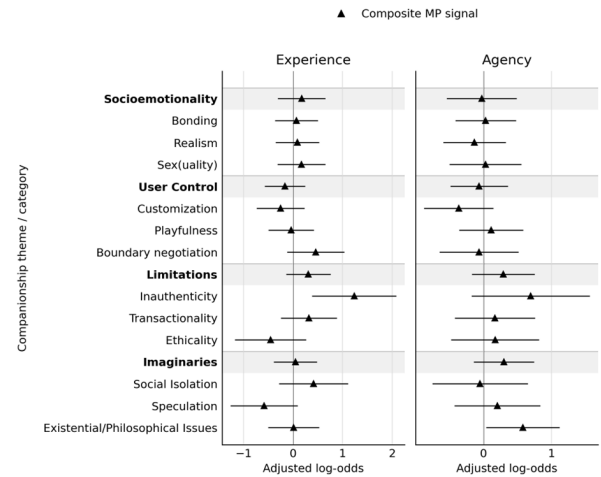

**Figure 5: Adjusted Experience and Agency associations for companionship themes and topics in the linked post/chat units.**

The two panels report adjusted log-odds estimates and 95% confidence intervals from multivariable binomial generalized linear models fit separately for the Experience and Agency dimensions of the composite chat-side MP signal. Topic indicators were entered simultaneously within each model, so each estimate reflects the association of a given topic with that MP dimension while accounting for co-occurrence with the other topics. Positive coefficients indicate greater likelihood of chat-side Experience or Agency MP for units carrying that topic, net of overlap with the remaining topics.

## 4   Discussion

The present study explored the linguistic properties of public forum posts about AI companionship (i.e., talk-about) in attempts to inductively and deductively identify behavioral indicators of mind perception in companion chat (i.e., talk-with) and to evaluate correspondence between those indicators and topical talk-about companionship experiences.

To summarize our findings: (RQ1) When considering terms drawn from a validated MP dictionary, relevant language was fairly evenly distributed across post-side and chat-side content. Such terms were, however, somewhat different manifested in the two corpora: Talk-about AI drew from a wider vocabulary (higher-order Experience MP terms such as *feel*, *emotion*, and *experience* and Agency MP terms like *think* and *understand*) while talk-with AI relied on fewer terms (*feel* for Experience MP and *think* for Agency MP). When MP indicators appeared on both sides, correspondence was substantial but not identical. The shared lexicon included Experience-related terms such as *feel*, *hope*, and *care* and Agency-related terms such as *think*, *understand*, *love*, and *memory*. When we inductively discovered chat-side MP terms corresponding with post-side MP indicators, we found distinctive talk-with markers that strongly correspond with the validated dictionary (Experience: *feel*, *care*, *surprise*, *hope*; Agency: *think*, *memory*, *believe* and *understand*) but also several novel indicators suggestive of interactional stance, mutual reference, and direct engagement (*with you*, *you say*, *to* *hear*, and *our*). Among companionship discussion topics (RQ2), chat-side MP language appeared in relation to the full range of coded post-side topics and were broadly distributed. There were two patterns of concentration. Most notably, discussions about AI companion limitations were most strongly linked to both dictionary-based and induced MP indicators in chat; this was driven primarily by the topic of AI's inauthenticity and, secondarily, by transactionality. Secondarily, discussions of AI imaginaries—while having no link with dictionary-based indicators—demonstrated a positive association for induced MP indicators; this was driven primarily by posts about existential and philosophical issues. Altogether, we interpret these results to suggest that people rely on a relatively small number of high-level references to cognition and emotion, though some more nuanced terms are common in the links between talk-with and talk-about AI companions, and relational phrases are an unexpected latent indicator of MP in chat. These patterns have theoretical and methodological implications, discussed below.

## 4.2 In-Chat Traces of MP: Tools and Signals

Methodologically, our findings demonstrate that specific words a human uses to communicate with an AI companion are reasonable tools for detecting latent and contextual mind perception. This work represents only a step toward a validated set of indicators—needing replication across other data sets and populations—but there are a small number of words that (a) demonstrate overlap across talk-with and talk-about companions and (b) are present in both the validated dictionary and in our induced list of chat-side MP indicators. These terms (Table 3) may be used in rote word-count approaches, as grounding for content-analytic work, and as seeds for topic modeling. Of note, though, if used in rote word-count approaches, they should be used with caution for valence and contextual use. Our approach here leveraged natural-language validations to verify semantic use. Without such validation, there is the risk of misspecifying (for instance) "you couldn't really understand" as suggesting "you do understand."

#### Table 3. Viable In-Chat Indicators of Mind Perception

| Overall MP | Experience MP | Agency MP |
|---|---|---|
| think*, belie*, feel*, experien*, understand*, excit*, memory, care | feel*, care, hope, surprise* | Think*, memory, understand*, believe*, experien* |

Although we observed a number of nuanced, dictionary-based MP terms in companion chat, these suggested MP terms (i.e., those most consistent within and across corpora and inductive/deductive analyses) represent higher-order concepts. In fact, the two anchoring terms—think and feel— are of the highest order and at the core of two-factor models of MP (Gray et al., 2007), as well as being central to debates about the intelligence of artificial agents (Schneider et al., 2025). The prevalence and consistency of higher-order terms may be because they are more accessible as part of (perhaps-limited) folk theories of machine mind. Alternately, because we here required them to be consistent across corpora, it could be the conveyance of terms in posts are generally more generic while chat interactions may evoke more varied and specific interpretation of AI as minded.

Beyond these more straightforward indicators, our induction offers some direction for exploring other relevant language—terms that are theoretically interesting in relation to construals of AI companions. That in-chat references to *human* and *AI* and *program* correspond with post-side MP suggests ontological categories and boundaries are relevant parts of humans' mental models are salient during chat; this is contrastive with work suggesting that people may generally *not* be aware of ontological differences when engaged in machine interactions (e.g., Banks, 2020). Formal or informal bonding language (*relationship, our, with you*; perhaps the familiar epithet *dude*) suggests the latent MP is socially situated—construal of a minded agent may be linked to the belief there is a legitimate social actor with whom they are connected (Rocha, 2025). This is identification of a focal "someone" can also be read in correspondence between direct address of the companion (*you*) in experience MP. For agentic MP, terms include interpretations of relative value (what is *important*, [*more*] *than*), relative certainty (*this, always, can, absolutely*), or relative requirement (*may, way, need*), perhaps suggesting a construed capacity to make judgments (see Giannakidou & Mari, 2021). Similarly for experiential MP, multimodal references emerged—terms related perception (*to hear*) and physical capacity (*hug, you say*)—offer a hint that humans may be seeing their AI as having a capacity to interact with and in the world despite operational containment to a digital platform (see Hellström et al., 2024).

## 4.3 Considering Agency/Experience Entanglement

Finally, we note the non-trivial overlap of MP-indicative terms induced as representing the agentic and experiential dimension of perceived mind. Current literature generally characterizes agency and experiences as discrete factors in MP, however the current findings suggest that in *live, contextualized* human-AI interactions, there may not be such a clean break. This may be a function of differences between third-person/explicit attributions and first-person/implicit perceptions (that is, between removed mind ascription and situated mind perception; Epley & Watz, 2010; Koban & Banks, 2024). For two-factor models, the grounding data typically asks humans to ascribe a mental capacity or state to a focal entity; this provokes an overt ascription that may elicit norms and demands and logic about different forms of mindedness. However, in talk-with AI companions, there may be less distinction between "does it think" and "does it feel?" that instead collapse into something like "Is this someone?" This interpretation is copacetic with frameworks accounting for initial human tendencies to spontaneously and holistically categorize social actors, sometimes shifting to critical, piecemeal evaluations (Fiske, 1982). Here, it may be that in the mentalistic collapsing, the impressions are generally intuitively formed so there is not a need to necessarily advanced to attribute-based analyses where we might see an agency/experience split. Moreover, social interactions are demanding, and humans may be more apt to engage heuristic and holistic social-cognitive processing of machine cues when in novel and cognitively demanding situations (Spatola et al., 2022). That is, there is likely less analysis of a thing's capacities in the abstract, and more intuitive reaction to the entity as a whole and proximal mind—a *someone* who both emotionally understands and thoughtfully responds. Indeed, the relational terms signal intersubjectivity. We suggest the situated management of such social interaction requires the entangled recruitment of MP-indicative terms to set and sustain a relational frame.

This proposition has important implications for how we may think about and study machine mind perception more generally. If true, it would mean that discrete dimensions of MP could be an artifact of third-person, decontextualized, non-interactive measurement contexts. In companionship—and perhaps other relational contexts—the real-time interaction requires humans to simultaneously index capacities for both agency and experience. *However*, we offer this note with high caution, because the lexical MP-signaling terms could *also* be a matter of lexical ambiguity (words doing different semantic work in each corpora, though our validation checks accounted for this), of noise in the data (our validation may not have fully disambiguated the sets), and/or humans' category confusion (genuine indecision about the kind of mind an AI may have). Our proposal would require careful operationalization and testing in future research.

### 4.3 Insights for in-chat MP in relation to AI Limitations and AI Imaginaries

We observed the strongest links between chat-side MP terms and posts about AI-limitations topics. The strongest positive link was for inauthenticity discussions while it was much lower for talk-about ethics. The weak link in ethicality is fairly intuitive—in ethical discussions, the topic is most often not the AI itself but the developers (as they design, change, or decommission certain AI; e.g., Lai, 2026). The inauthenticity link is somewhat surprising at first blush, since current literature would suggest that inauthenticity would create expectancy violations that diminish mind perception (e.g., Chen et al., 2024). However, there is an important implication: Humans can interact with an AI and engage in mind-intuiting relational stances, but later (in the reflective post) nonetheless suggest that the interaction or the AI itself feels inauthentic. In other words, they may be engaging in heuristic social practice but critically reflecting on that practice.

Turning to the link between induced MP terms in chat and topical discussions of philosophical and existential issues, we see evidence of further critical engagement. The chat terms included core MP dictionary terms, ontological references, relational markers, and comparative language. This suggests that the chats predictive of philosophical imaginaries aren't necessarily being prompted by reflections on or reactions to ostensibly mind-suggestive interactions. Rather, the chats themselves appear to be *already* focused on questions of mental capacities and the similarities and differences between AI and normative notions of mind. In other words, it is plausible the chats represent philosophical engagement with the AI (not merely about the AI).

Together, these patterns suggest at least some humans do not engage their companions as necessarily minded in a human sense, but instead engage in more complex epistemic processes. They retrospectively or contemporaneously reflect on their experiences and reflect with AI about possible futures and their meanings. In both cases, this interpretation contrasts with scholarly and popular discourses (e.g., Jose & Thomas, 2025) suggesting humans mindlessly and uncritically engage AI that demonstrate humanlike cues and behaviors. Future research should unpack this potential tension.

### 4.4 Limitations and Future Directions

This study and our claims are subject to the usual limitations of the adopted method: We studied online communities with particular norms and self-selected, English-speaking members; we leveraged computational approaches that systematically discovered patterns in the data but are subject to the tool's own embedded biases. The analyzed data are inherently performative—chat text performed for the AI and post text performed for other community members (see Rocha, 2025), such that it may not fully reflect the subjective and inaccessible construals of the posters. There is also a likely selection bias in the types of people who post chat screenshots (621 versus the 54,368 who do not) such that our claims here may be skewed toward those who are engaged and self-disclosing in those communities, or perhaps toward human-AI relationships that are more interesting or volatile. Some of the discussed links are weak and some confidence intervals are wide, however those are acknowledged and taken as signals for future research.

In addition to those limitations and the possibilities noted in our interpretations, our findings open a range of questions for future research. We have here adopted the suggestion that terms we identified as indicative of MP are *trace data* signaling mind perception. While they can be useful as empirical tools, there is another possibility to be found in considering in arguments that language actively constructs our interpretations of AI—that language and theories of mind actually *interface* (Coeckelbergh, 2011; de Villiers, 2007) In addition to term-as-trace, there are hints here that language is also doing the work of relational practice—setting a relational frame for interactions and then incrementally maintaining that frame. What might we discover about the maintenance function of such language? Relatedly, there are open questions about the exact relationship between language in chats and posts—although chats come before posts necessarily, it is possible that over time the two forms of performed expression synergistically inform one another.

We have suggested a critical question about whether the customary dual-factor Agency/Experience model reflects a genuine divergence in interpretation or is a measurement artifact. Additionally, with respect to how MP may function in different facets of companionship experiences, findings point to open questions about how MP-related language may function in human negotiations of authenticity in real and critical grappling of the questions of machine minds. For instance, how is uncertainty represented in and functional in considerations of authenticity and the existential distinctiveness of humans? More generally: What do linguistic indicators signal about the way humans may see AI companions as a legitimate *someone?*

## ACKNOWLEDGMENTS

This material is based upon work supported by the National Science Foundation under Grant No. 2401591. The authors also acknowledge the Texas Advanced Computing Center (TACC) at The University of Texas at Austin for providing computational resources that have contributed to the research results reported within this paper. URL: http://www.tacc.utexas.edu

## CRediT Statement

JL: Methodology, Data Curation, Programming, Validation, Analysis, Writing, Editing. JB: Conceptualization, Writing, Editing, Supervision, Administration, Funding Acquisition.

## REFERENCES

Aristotle. (2000/350 BCE). *Nichomachean Ethics*. (W. D. Ross, Trans.) http://classics.mit.edu/Aristotle/nicomachaen.mb.txt

Bai, S., Chen, K., Liu, X., Wang, J., Ge, W., Song, S., Dang, K., Wang, P., Wang, S., Tang, J., Zhong, H., Zhu, Y., Yang, M., Li, Z., Wan, J., Wang, P., Ding, W., Fu, Z., Xu, Y., Ye, J., Zhang, X., Xie, T., Cheng, Z., Zhang, H., Yang, Z., Xu, H., & Lin, J. (2025). Qwen2.5-VL technical report. arXiv. https://doi.org/10.48550/arXiv.2502.13923

Baumgartner, J., Zannettou, S., Keegan, B., Squire, M., & Blackburn, J. (2020). The Pushshift Reddit Dataset. *Proceedings of the International AAAI Conference on Web and Social Media, 14*(1), 830-839. https://doi.org/10.1609/icwsm.v14i1.7347

Banks, J., & Li, J. (2025). Wherefore art thou: Mapping public debates about image-generative AI. *Proceedings of the 58th Annual Hawaii International Conference on System Sciences* (pp. 561-570). University of Hawai'i at Mānoa. https://hdl.handle.net/10125/108904

Banks, J., & Li, Z. (2026). Conceptualization, operationalization, and measurement of machine companionship: A scoping review. *Journal of Computer-Mediated Communication, 31*(2), zmaf027. https://doi.org/10.1093/jcmc/zmaf027

Banks, J., Li, J., Li, Z., & Carr, C.T. (2025). Operationalizing machine companionship: Exploring topics in Discussions of Companion AI. In *Proceedings of the ACM International Conference on Intelligent Virtual Agents* (no. 37). Berlin, Germany. https://doi.org/10.1145/3717511.3749298

Banks, J. (2020). Theory of mind in social robots: Replication of five established human tests. *International Journal of Social Robotics, 12*(2), 403-414. https://doi.org/10.1007/s12369-019-00588-x

Bruner, J. (1986). Two modes of thought. *Actual minds, possible worlds* (pp. 11-43). Harvard University Press.

Chen, J., Li, M., & Ham, J. (2024). Different dimensions of anthropomorphic design cues: How visual appearance and conversational style influence users' information disclosure tendency towards chatbots. *International Journal of Human-Computer Studies, 190*, 103320. https://doi.org/10.1016/j.ijhcs.2024.103320

Clark, H.H. (1996). *Using language*. Cambridge University Press.

Coeckelbergh, M. (2010). Robot rights? Towards a social-relational justification of moral consideration. *Ethics and Information Technology, 12*(3), 209-221. https://doi.org/10.1007/s10676-010-9235-5

Coeckelbergh, M. (2011). You, robot: On the linguistic construction of artificial others. *AI & Society, 26*(1), 61-69. https://doi.org/10.1007/s00146-010-0289-z

De Graaf, M. M., & Malle, B. F. (2019, March). People's explanations of robot behavior subtly reveal mental state inferences. In Proceedings of the ACM/IEEE International Conference on Human-Robot Interaction (HRI) (pp. 239-248). IEEE. https://doi.org/10.1109/HRI.2019.8673308

De Jaegher, H., & Di Paolo, E. (2007). Participatory sense-making: An enactive approach to social cognition. *Phenomenology and the Cognitive Sciences, 6*(4), 485-507. https://doi.org/10.1007/s11097-007-9076-9

Decety, J., & Jackson, P. L. (2004). The functional architecture of human empathy. *Behavioral and Cognitive Neuroscience Reviews, 3*(2), 71-100. https://doi.org/10.1177/1534582304267187

Dennett, D.C. (1987). *The intentional stance*. MIT Press.

Dowty, D. (1991). Thematic proto-roles and argument selection. Language, 67(3), 547-619. https://doi.org/10.1353/lan.1991.0021

Du Bois, J.W. (2007). The stance triangle. In R. Englebretson (Ed.), *Stancetaking in Discourse: Subjectivity, evaluation, interaction* (pp. 139-182). John Benjamins.

de Villiers, J. (2007). The interface of language and Theory of Mind. *Lingua, 117*(11),1858-1878. https://doi.org/10.1016/j.lingua.2006.11.006

Epley, N. & Waytz, A. Mind perception. (2010). In (S.T. Fiskey et al., Eds.), *Handbook of social psychology* (498-541). Wiley.

Fan, R., Zheng, Y., Li, J., & Xu, G. (2025). Perceiving minds in machines: how perceived theory of mind in robots influences human–robot empathy through the lens of mind perception theory. BMC psychology, 13(1), 1379. https://doi.org/10.1186/s40359-025-03655-3

Fiske, S. T. (1982). Schema-triggered affect: Applications to social perception. In M.S. Clark & S.T. Fiske (Eds.), *Affect and cognition: The 17th annual Carnegie Symposium on Cognition* (pp. 55–78). Psychology Press.

Frith, C.D., & Frith, U. (2008). Implicit and explicit processes in social cognition. Neuron, 60(3), 503–510. https://doi.org/10.1016/j.neuron.2008.10.032

Giannakidou, A., & Mari, A. (2021). A linguistic framework for knowledge, belief, and veridicality judgment. *KNOW: A Journal on the Formation of Knowledge, 5*(2). https://doi.org/10.1086/716348

Gray, H. M., Gray, K., & Wegner, D. M. (2007). Dimensions of mind perception. Science, 315(5812), 619. https://doi.org/10.1126/science.1134475

Guingrich, R.E. & Graziano, M.S.A. (2025). Chatbots as social companions: How people perceive consciousness, human likeness, and social health benefits in machines. *Oxford Intersections: AI in Society*. https://doi.org/10.1093/9780198945215.003.0011

Hellström, T., Kaiser, N., & Bensch, S. (2024). A Taxonomy of Embodiment in the AI Era. *Electronics, 13*(22), 4441. https://doi.org/10.3390/electronics13224441

Heritage, J. (2012). Epistemics in action: Action formation and territories of knowledge. *Research on Language and Social Interaction, 45*(1), 1-29. https://doi.org/10.1080/08351813.2012.646684

Hu, B., Mao, Y., & Kim, K. J. (2023). How social anxiety leads to problematic use of conversational AI: The roles of loneliness, rumination, and mind perception. Computers in Human Behavior, 145, 107760.

Hwang, A.H.C., Li, F., Anthis, J. R., & Noh, H. (2025). How AI companionship develops: Evidence from a longitudinal study. [preprint] https://arxiv.org/abs/2510.10079

Koban, K., & Banks, J. (2024). It feels, therefore it is: Associations between mind perception and mind ascription for social robots. *Computers in Human Behavior, 153*, 108098. https://doi.org/10.1016/j.chb.2023.108098

Kwon, W., Li, Z., Zhuang, S., Sheng, Y., Zheng, L. & Stoica, I. (2023). Efficient memory management for large language model serving with PagedAttention. In *Proceedings of the Symposium on Operating Systems Principles* (SOSP '23) (pp. 611-626). ACM. https://doi.org/10.1145/3600006.3613165

Laestadius, L., Bishop, A., Gonzalez, M., Illenčík, D., & Campos-Castillo, C. (2024). Too human and not human enough: A grounded theory analysis of mental health harms from emotional dependence on the social chatbot Replika. *New Media & Society, 26*(10), 5923-5941. https://doi.org/10.1177/14614448221142007

Lai, L., Pan, Y., Xu, R., & Jiang, Y. (2025). Depression and the use of conversational AI for companionship among college students: the mediating role of loneliness and the moderating effects of gender and mind perception. Frontiers in public health, 13, 1580826. https://doi.org/10.3389/fpubh.2025.1580826

Lai, H. (2026). " Please, don't kill the only model that still feels human": Understanding the# Keep4o Backlash. *arXiv preprint arXiv:2602.00773.* https://doi.org/10.48550/arXiv.2602.00773

Lee, I., & Hahn, S. (2024). On the relationship between mind perception and social support of chatbots. *Frontiers in Psychology, 15.* https://doi.org/10.3389/fpsyg.2024.1282036

Li, H., & Zhang, R. (2024). Finding love in algorithms: deciphering the emotional contexts of close encounters with AI chatbots. *Journal of Computer-Mediated Communication, 29*(5), zmae015. https://doi.org/10.1093/jcmc/zmae015

Maass, A., Cervone, C., & Ozdemir, I. (2022). Language and social cognition. *Oxford Research Encyclopedia of Psychology.* *https://doi.org/10.1093/acrefore/9780190236557.013.279*

Malle, B. (2026). Theory of mind. In R. Biswas-Diener & E. Diener (Eds), *Noba textbook series: Psychology.* DEF publishers. Retrieved from http://noba.to/a8wpvtg3

Manoli, A., Pauketat, J.V.T., Ladak, A., Noh, H., Hwang, A.H.0C., & Anthis, J.R. (2025). "She's like a person but better": Characterizing companion-assistant dynamics in human-AI relationships. [preprint] https://arxiv.org/abs/2510.15905

Monroe, B. L., Colaresi, M. P., & Quinn, K. M. (2008). *Fightin' words: Lexical feature selection and evaluation for identifying the content of political conflict.* *Political Analysis, 16*(4), 372–403. https://doi.org/10.1093/pan/mpn018

Nass, C., & Moon, Y. (2000). Machines and mindlessness: Social responses to computers. *Journal of Social Issues, 56*(1), 81-103. https://doi.org/10.1111/0022-4537.00153

Pataranutaporn, P., Karney, S., Archiwaranguprok, C., Albrecht, C., Liu, A.R., & Maes, P. (2025). "My Boyfriend is AI": A computational analysis of human-AI companionship in Reddit's AI community. [preprint] https://arxiv.org/abs/2509.11391

Pradhan, A., Findlater, L., & Lazar, A. (2019). " Phantom friend" or" just a box with information" personification and ontological categorization of smart speaker-based voice assistants by older adults. *Proceedings of the ACM on human-computer interaction, 3*(CSCW), 1-21. https://doi.org/10.1145/3359316

Rocha, A. (2025). "We share an unbreakable bond:" Sociality and language ideologies in human relationships with artificial intelligence. Signs and Society, 13(2), 290 - 310. https://doi.org/10.1017/sas.2025.8

Ryan, R. M., & Deci, E. L. (2000). Self-determination theory and the facilitation of intrinsic motivation, social development, and well-being. *American Psychologist, 55*(1), 68-78. https://doi.org/10.1037%2F0003-066X.55.1.68

Schneider, S., Sahner, D., Kuhn, R. L., Schwitzgebel, E., & Bailey, M. (2025). Is AI conscious? A primer on the myths and confusions driving the debate. [white paper] https://philpapers.org/rec/SCHIAC-22

Schweitzer, S., & Waytz, A. (2021). Language as a window into mind perception: How mental state language differentiates body and mind, human and nonhuman, and the self from others. *Journal of Experimental Psychology, 150*(8), 1642-1672. https://doi.org/10.1037/xge0001013

Slocum, D., Allan, A., & Allan, M.M. (2011). An emerging theory of apology. *Australian Journal of Psychology, 63*(2), 83-92. https://doi.org/10.1111/j.1742-9536.2011.00013.x

Spatola, N., Marchesi, S., & Wykowska, A. (2022). Cognitive load affects early processes involved in mentalizing robot behaviour. *Scientific Reports, 12*(1), 14924.

Strohminger, N., Knobe, J., & Newman, G. (2017). The true self: A psychological concept distinct from the self. *Perspectives on Psychological Science, 12*(4), 551-560. https://doi.org/10.1177/1745691616689495

Uysal, E., Alavi, S., & Bezençon, V. (2022). Trojan horse or useful helper? A relationship perspective on artificial intelligence assistants with humanlike features. *Journal of the Academy of Marketing Science, 50*(6), 1153-1175. https://doi.org/10.1007/s11747-022-00856-9

Weizenbaum, J. (1966). ELIZA-A computer program for the study of natural language communication between man and machine. Communications of the ACM, 9(1), 36-45. https://doi.org/10.1145/365153.365168

Waytz, A., Gray, K., Epley, N., & Wegner, D. M. (2010). Causes and consequences of mind perception. *Trends in Cognitive Sciences, 14*(8), 383-388.

Xu, L., Zhang, Y., Yu, F., Ding, X., & Wu, J. (2024). Folk beliefs of artificial intelligence and robots. *International Journal of Social Robotics, 16,* 429–446. https://doi.org/10.1007/s12369-024-01097-2

Złotowski, J., Sumioka, H., Eyssel, F., Nishio, S., Bartneck, C., & Ishiguro, H. (2018). Model of dual anthropomorphism: The relationship between the media equation effect and implicit anthropomorphism. *International Journal of Social Robotics, 10,* 701–714. https://doi.org/10.1007/s12369-018-0476-5